\input{aipcheck}

\documentclass[
    ,final            % use final for the camera ready runs
  ]
  {aipproc}

\layoutstyle{6x9}

\begin{document}

\title{Coupled-Channels Approach for Dissipative Quantum Dynamics in Near-Barrier Collisions}

\classification{24.10.-i, 24.10.Eq, 25.70.Jj, 25.60.Pj, 03.65.Yz, 05.30.-d}
\keywords{Open quantum system, Dissipation, Decoherence, Lindblad equation, Coupled channels, Quantum tunneling, Fusion, Deep-inelastic collisions}

\author{A. Diaz-Torres}{
  address={Department of Nuclear Physics, Research School of
Physical Sciences and Engineering, Australian National University,
Canberra, ACT 0200, Australia}
}

\author{D.J. Hinde}{
  address={Department of Nuclear Physics, Research School of
Physical Sciences and Engineering, Australian National University,
Canberra, ACT 0200, Australia}
}

\author{M. Dasgupta}{
  address={Department of Nuclear Physics, Research School of
Physical Sciences and Engineering, Australian National University,
Canberra, ACT 0200, Australia}
%%  ,altaddress={<author1 address>} % additional visiting address
}

\author{G.J. Milburn}{
  address={Centre for Quantum Computer Technology, University of
Queensland, St. Lucia, Queensland 4072, Australia}
}

\author{J.A. Tostevin}{
  address={Department of Physics, Faculty of Engineering and
Physical Sciences, University of Surrey, Guildford, Surrey GU2 7XH,
United Kingdom}
}

\begin{abstract}
 A novel quantum dynamical model based on the dissipative quantum dynamics of 
open quantum systems is presented. It allows the treatment of both deep-inelastic 
processes and quantum tunneling (fusion) within a fully quantum mechanical 
coupled-channels approach. Model calculations show the transition from pure state 
(coherent) to mixed state (decoherent and dissipative) dynamics during a near-barrier 
nuclear collision. Energy dissipation, due to irreversible decay of giant-dipole 
excitations of the interacting nuclei, results in hindrance of quantum tunneling.   
\end{abstract}

\maketitle

%%%%%%%%%%%%%%%%%%%%%%%%%%%%%%%%%%%%%%%%%%%%
%% MAINMATTER
%%%%%%%%%%%%%%%%%%%%%%%%%%%%%%%%%%%%%%%%%%%%

\section{INTRODUCTION}

Stationary state coupled-channels approaches have been very
successful \cite{Thompson} in explaining several collision
observables. However, there are still unsolved problems. For instance, 
the inability to describe elastic scattering and fusion measurements 
simultaneously \cite{Newton,Angeli} and, related, the more recent failure to 
describe in a physically consistent way the below-barrier quantum 
tunneling and above-barrier fusion yields \cite{Nanda2}.

These problems may be caused by the neglect of important physical 
processes (e.g., deep-inelastic) which cannot be treated within 
(standard) coupled-channels models. 
Measurements have shown that deep-inelastic 
processes occur even at sub-barrier incident energies \cite{Wolfs}, in competition 
with the process of quantum tunneling, and 
thus fusion. Energy dissipation associated with the deep-inelastic mechanism 
could thus play a significant role in the inhibition of tunneling at deep sub-barrier 
energies. This can also change the yield of direct reaction 
processes, including elastic and quasi-elastic channels. 

The understanding of this complex interplay, at near- and below barrier energies, requires a dynamical model which can describe coupling assisted tunneling with dissipation. Neither existing models of fusion nor of deep-inelastic scattering can address both energy dissipation and quantum tunneling. Quantum mechanical coupled-channels models describe tunneling without energy dissipation \cite{Thompson,Hagino}, whilst approaches to direct damped collisions treat the relative motion of the nuclei classically \cite{Schroeder}. 

We here report on a novel coupled-channels density matrix approach \cite{Alexis1} that overcomes these difficulties. We exploit the Lindblad axiomatic approach \cite{Lindblad1,Sandulescu} for open quantum systems, which is based on the concept of quantum semigroups and completely positive mappings. We refer to Ref. \cite{Alexis1} for a discussion of the suitability of the Lindblad theory for the treatment of low-energy collision dynamics. The coupled-channels description is formulated with Lindblad's equation for a {\em reduced} density matrix \cite{Lindblad1,Sandulescu}. It describes the dynamical evolution of the reduced system (comprising the relative motion of the nuclei plus selected, intrinsic collective excitations) that irreversibly interacts with two (model) ``environments''. Firstly, an environment inside the Coulomb barrier, which is related to the complexity of compound nucleus states. Secondly, one with a long range, associated with decay out of short lived (compared to the reaction time) internal vibrational states, e.g. the giant dipole resonance (GDR) of the colliding nuclei. Model calculations show that damping of the GDR results in quantum decoherence and energy loss as the nuclei overlap, inhibiting tunneling, and thus fusion.     

\section{COUPLED CHANNELS DENSITY MATRIX APPROACH}

\begin{figure}
  \includegraphics[height=.53\textheight]{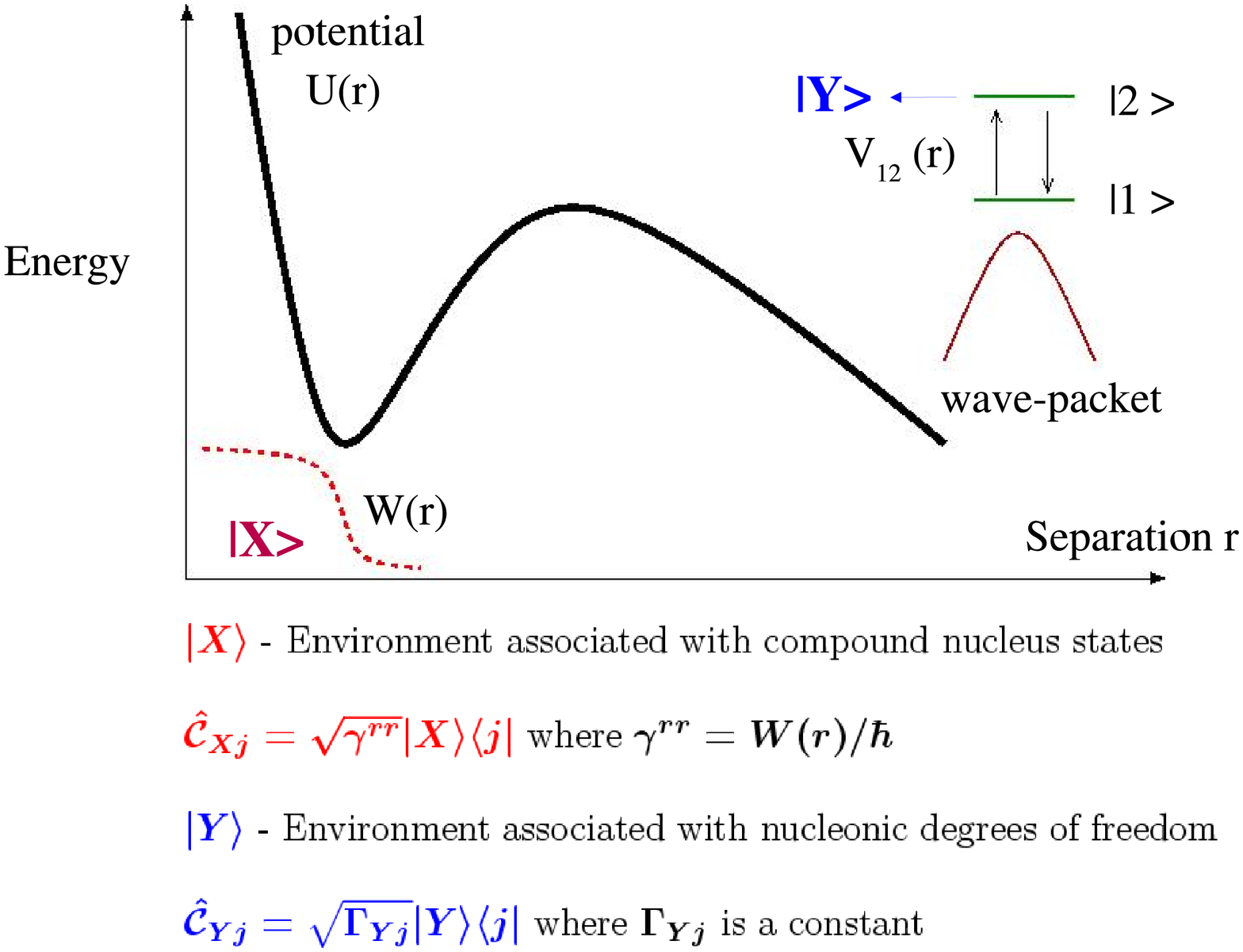}
  \caption{(Color online) Schematic picture of the model scenario. 
$\hat{\mathcal {C}}_{Xj}$ and $\hat{\mathcal {C}}_{Yj}$ are physically motivated 
Lindblad's operators for the two dissipative channels $|X \rangle$ and $|Y \rangle$, respectively. See text for further details.}
\label{Fig1}
\end{figure}

The model calculations are carried out using the scenario presented in 
Figure \ref{Fig1}. The basis will comprise two
asymptotic states (coupled-channels) $|1 \rangle$ and $|2 \rangle$. 
Channel $|1 \rangle$ is the (ground
states) entrance channel and is coupled to an inelastic state $|2
\rangle$ by a coupling interaction $V_{12}$. Two distinct sources of
irreversibility are also considered, modelled by two auxiliary
(environment) states $|X \rangle$ and $|Y \rangle$. The first
environmental coupling describes capture by the potential pocket
inside the fusion barrier. This simulates the irreversible and
dissipative excitations associated with the evolution from the two
separate nuclei to a compound nuclear system. In a stationary states
approach this loss of flux is approximated by imposing an imaginary
potential $-iW(r), W(r)>0,$ or an ingoing-wave boundary condition at
distances well inside the barrier. Here, these transitions are
described by an auxiliary state $|X \rangle$, to which \emph{all}
other states $|j \rangle $ couple, modelled \cite{Irene} by a
Lindblad operator $\hat{\mathcal {C}}_{Xj}=\sqrt{\gamma ^{rr}} |X
\rangle \langle j|$. The absorption rate to state $|X \rangle$ is
given by $\gamma^{rr} = W(r)/\hbar$ where $W(r)$ is taken as a Fermi
function with depth 10 MeV and diffuseness 0.1 fm, located at the
pocket radius of the nucleus-nucleus potential, $\approx 7$ fm. This
choice guarantees complete absorption inside the pocket. The fusion
probability is defined as the probability accumulating in this state
$|X \rangle $.

The second environment, whose explicit treatment will be seen to be
the most significant at lower energies, is associated with the
irreversible decay out of intrinsic excitations of the colliding
nuclei. Such decays are independent of the dynamical couplings.
Specifically, we will associate the only excited coupled channel
state $|2 \rangle$ with the GDR excitation. We then introduce a
second auxiliary state $|Y \rangle$, representing the bath of states 
in which the GDR is embedded, and to which only the GDR excitation
$|2 \rangle$ is coupled.

Thus, $|Y \rangle$ and/or $|X \rangle$ supplement the two intrinsic
states $|1 \rangle$ and $|2 \rangle$ that comprise the two coupled
channels. Both of the auxiliary states refer to complex excitation
modes of the nuclei, associated with nucleonic degrees of freedom
and compound nucleus states, respectively. They provide intuitive
and formal channels \cite{Irene} for describing irreversible
coupling and loss of probability from the system to these
environments, couplings that enter \emph{only} through 
Lindblad's dissipative Liouvillian \cite{Alexis1}.
$|Y \rangle$ is also assumed to couple to $|X \rangle$ at the
appropriate range of separations. Probability accumulating in state
$|Y \rangle$ outside of this $|X \rangle$ pocket may be identified
with deep-inelastic processes.

Dynamical calculations proceed as follows. The reduced density matrix results from 
the representation of the reduced density operator $\hat{\rho}(t)$ in an asymptotic (product) basis of coordinate states and intrinsic energy states of the individual 
nuclei. Its dynamical evolution is dictated by a set of Lindblad's coupled 
equations \cite{Alexis1} for the matrix elements, which defines a time-dependent 
coupled-channels problem with an initial value. The initial pure-state density matrix, describing the well-separated nuclei (in their ground states) with a wave-packet in
their relative motion, has ${\rm Tr} [\hat{\rho}^2] = 1$. The purity of
this state, conserved under unitary time evolution, is destroyed 
(${\rm Tr}[\hat{\rho}^2] < 1$) if the environment causes a loss of 
quantum coherence. The {\em degree of decoherence} is quantified by a loss of
purity of the density matrix. Having
solved the time-evolution of the density-matrix numerically \cite{Alexis1}, the
expectation value of reaction observables $\hat{\mathcal O}$ is
given by evaluating the appropriate trace $\langle \hat{\mathcal
O}(t) \rangle = {\rm Tr}[\hat{\mathcal O}\hat{\rho}(t)]$. In this way, 
probabilities of reaction channels can be determined.

We have performed test calculations for a head-on collision of $^{16}$O + $^{144}$Sm at 
energies below the nominal Coulomb barrier ($V_B = 61.1$ MeV). The Hamiltonian of 
the reduced system is the same as that in the fusion model implemented in 
{\sc ccfull} \cite{Hagino}. 
The form of the bare nuclear potential between the two nuclei, consistent with
the stated $V_B$, is a Woods-Saxon potential with ($V_0, r_0,a_0)
\equiv$ ($-$105.1 MeV, 1.1 fm, 0.75 fm). The Coulomb potential was
that for two point charges. The $^{16}$O projectile was taken to be
inert and the $^{144}$Sm target was allowed to be excited to a GDR
vibrational state. The dynamical nuclear coupling of the ground
state $|1 \rangle$ to the vibrational state $|2 \rangle$, with
excitation energy $E_{1^{-}}=15$ MeV, has a macroscopic deformed
Woods-Saxon form with a deformation parameter of $\beta_1=0.2$.

The time step for the density-matrix propagation was $\Delta t =
10^{-22}$ s, and the radial grid ($r=0-250$ fm) was evenly spaced
with $M=512$ points. The relative motion of the two nuclei in the
entrance channel $|1 \rangle$ was described by a minimal-uncertainty
Gaussian wave-packet, with width $\sigma_0 = 20$ fm, initially
centered at $r=150$ fm, and was boosted towards the target with the
appropriate average kinetic energy for the entrance channel energy
$E_0$ required. The FWHM energy spread of the wave-packet is $\sim 3
\%$. The numerical accuracy of the time evolution was checked using
a fully coherent, time-dependent calculation, excluding coupling to
states $|X \rangle $ and $|Y \rangle $. It was confirmed that the
normalisation and purity of the density-matrix, Tr$[\hat{\rho}]$ =
Tr$[\hat{\rho}^2]=1$, and the expectation value of the system energy
Tr$[\hat{H} \hat{\rho}]$ were maintained with high accuracy over the
required number of time steps, typically 700 for the full duration
of the collision.

\begin{figure}
%\begin{center}
%\begin{tabular}{cc}
\includegraphics[height=.5\textheight]{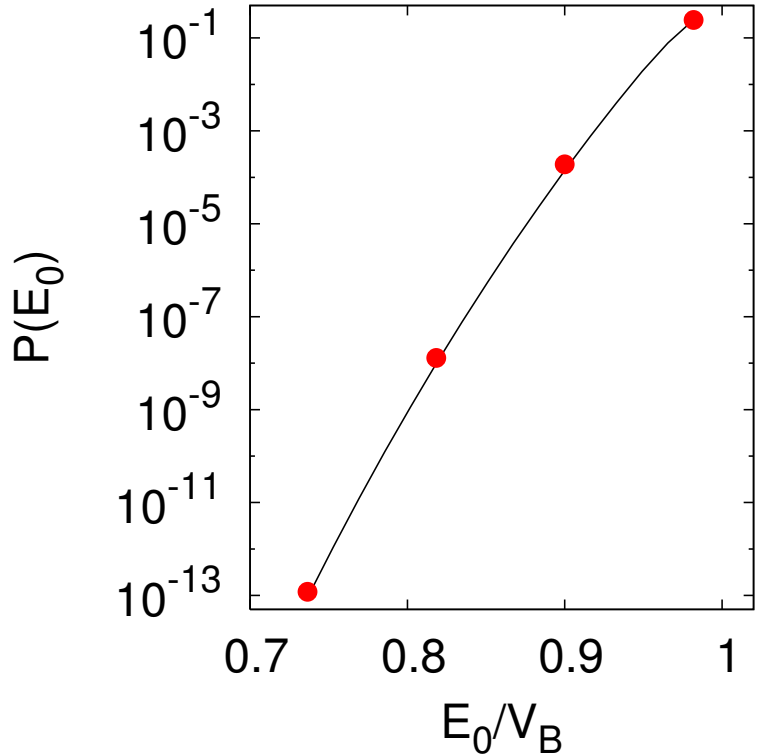}
%\end{tabular}
\caption{(Color online) The energy dependence of the s-wave
tunneling probability calculated with the density matrix (solid
points) and the coupled-channels {\sc ccfull} methods (full line). 
See text for further details.
}
\label{Figx}
%\end{center}
\end{figure}

The importance of the two, spatially distinct, sources of
environment couplings were studied. Calculations were first
performed only including the effect of coupling of the intrinsic 
coupled channels $|1 \rangle$ and $|2 \rangle$ to the capture state 
$|X \rangle $. Calculations were carried out for $E_0$ = 45, 50, 55 and 
60 MeV incident energy. The calculated state purity Tr$[\hat{\rho}^2]$ and
the energy dissipation Tr$[\hat{H} (\hat{\rho_0}- \hat{\rho})]$ post
the collision (after 700 time steps) are shown in the left panels in
Table \ref{Table}. For sufficiently sub-barrier energies, $E_0\leq
55$ MeV, it is evident that time-evolution in the presence of state
$|X \rangle$ essentially maintains coherence and is non-dissipative.
There is however loss of purity and dissipation at the highest
energy. It is interesting therefore to compare the density-matrix
and the predictions of {\sc ccfull} (that uses 
stationary Schr\"odinger's dynamics with an ingoing wave boundary 
condition). 
This is done here only for
calculations of the tunneling probability $P(E_0)$, in a relative
s-wave, shown in Figure \ref{Figx}. These comparisons, of
necessity, require convolution of the $\ell=0$ partial wave
penetrabilities $T_0(E)$ from {\sc ccfull} with the energy
distribution $f(E,E_0)$ of the chosen initial wave packet. That is,
$P(E_0) \equiv \int dE\, f(E,E_0)\, T_0(E)$. The $P(E_0)$, shown as
a function of $E_0/V_B$ in Figure \ref{Figx}, are in very good
agreement showing the appropriateness of stationary state
coupled-channels calculations for this observable within the
dynamical scheme of states $|1 \rangle$, $|2 \rangle$ and $|X
\rangle$. It is our contention that the dissipation associated with
state $|X \rangle$, while significant at 60 MeV, is strongly
localised inside the barrier and thus does not impact upon the
barrier penetrability. We will now show that the same is not true
for the more spatially-extended dissipation due to the GDR decay
environment $|Y \rangle$.

\begin{table} \caption{The calculated density matrix purity
Tr$[\hat{\rho}^2]$ and energy loss $\Delta E =$Tr$[ \hat{H}
(\hat{\rho_0}- \hat{\rho})]$ following time-evolution (for 700 time
steps) when including only the state $|X \rangle$ (left entries) and
both states $|X \rangle $ and $|Y \rangle$ (right entries)
environmental couplings. The GDR coupling strength used was
$\beta_1=0.2$. } \label{Table}
\medskip
%\begin{center}
\begin{tabular}{c|cc|cc}
\hline &\multicolumn{2}{c}{State $|X \rangle $} &
\multicolumn{2}{|c}{States $|X \rangle $ and $|Y\rangle$}
\\
\hline
$E_0$ (MeV) &~~~Tr$[\hat{\rho}^2]$~~&~~$\Delta E$ (MeV)~~ &~~ Tr$[\hat{\rho}^2]$~~&~~$\Delta E $ (MeV)~\\
\hline
45 &1.0000&0.0004&0.9196&1.8718\\
50 &1.0000&0.0004&0.8977&2.6744\\
55 &0.9996&0.0109&0.8759&3.6100\\
60 &0.6067&14.862&0.5127&18.908\\
\hline \end{tabular}
%\end{center}
\end{table}

The treatment of the irreversible GDR decay (with a spreading width
of 6 MeV) to the bath of surrounding complex states (represented by
$|Y \rangle$) was included by switching on the coupling of the
intrinsic inelastic state $|2 \rangle$ to $|Y \rangle$.
Unlike the coupling to $|X \rangle$, a major part of the inelastic
excitation of the system gives access to $|Y \rangle$  before the
wave packet encounters the fusion barrier. The onset of decoherence,
the purity of the density matrix, and the associated energy
dissipation are shown in the right hand entries in Table~\ref{Table}.

\begin{figure}[h]
%\begin{center}
\includegraphics[height=.5\textheight]{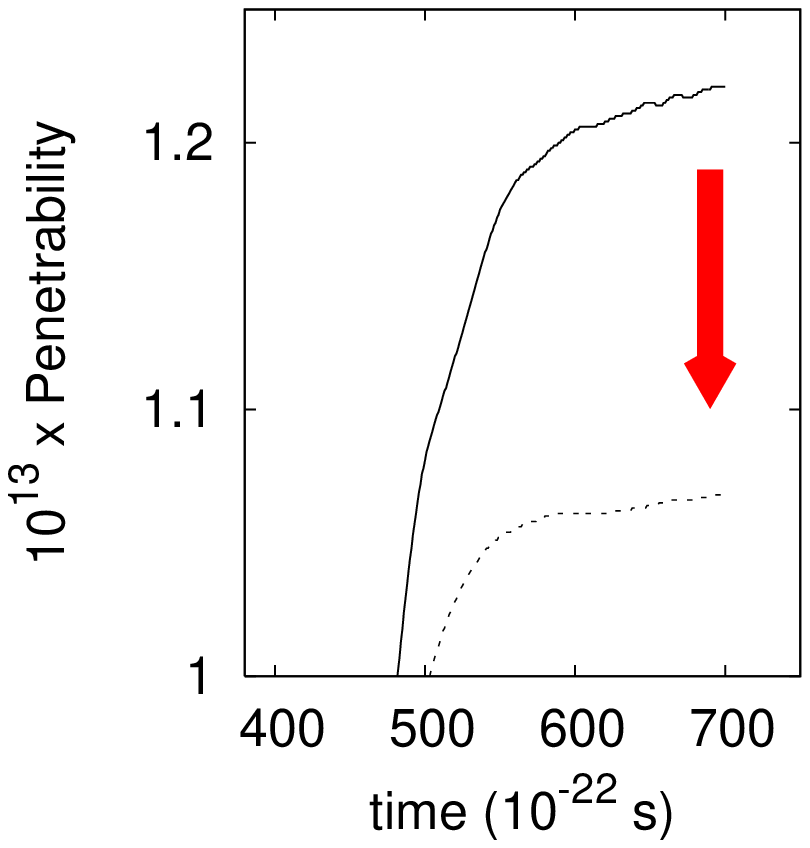}
\caption{(Color online) Time-dependence of the probability
trapped in $|X \rangle$ for $E_0=45$ MeV. The full curve includes
states $|1 \rangle$, $|2 \rangle$ and $|X \rangle$. The dotted curve
adds the irreversible decay of $|2 \rangle $ to $|Y \rangle $. The
calculations are for $\beta_1=0.2$.} \label{Figure3}
%\end{center}
\end{figure}

\begin{figure}[h]
%\begin{center}
\includegraphics[height=.5\textheight]{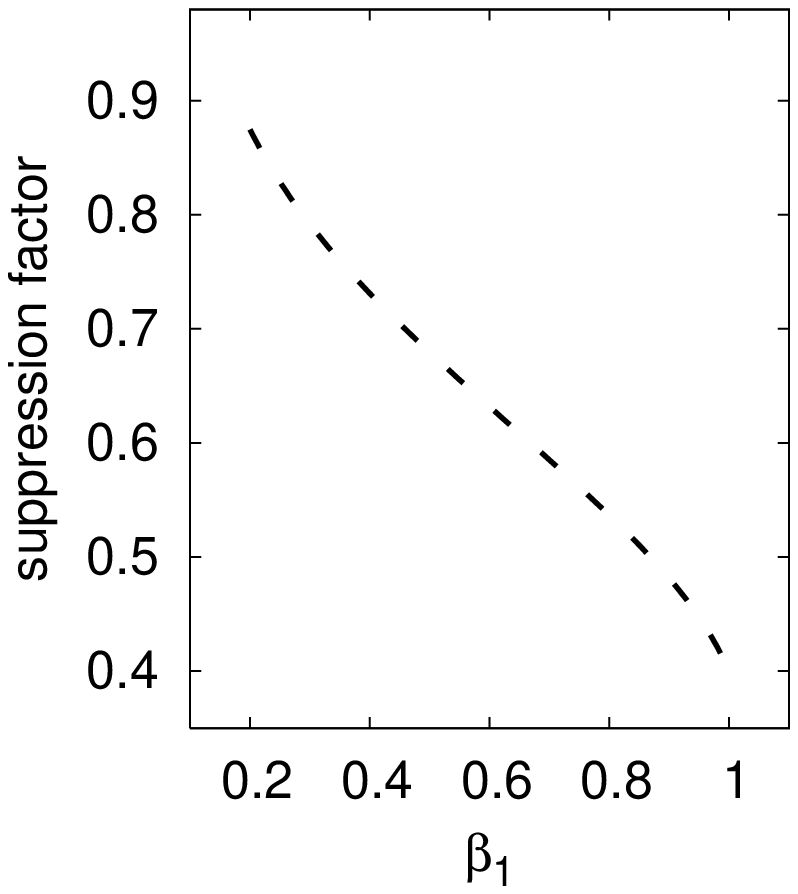}
\caption{Calculated suppression of
the probability trapped in $|X \rangle$ as a function of the assumed
$\beta_1$ value for $E_0=45$ MeV.} \label{Figure4}
%\end{center}
\end{figure}

The probability trapped under the fusion barrier is associated with
GDR collective vibrational energy being irreversibly removed from
the coherent dynamics into surrounding bath states (heat). This is then
no longer available for relative motion, or tunneling. Such energy
loss can be correlated with deep inelastic processes, seen
experimentally, that compete with fusion in reactions involving
heavy nuclei \cite{Wolfs}.

Figure\ \ref{Figure3} shows the time evolution of the probability
trapped in the potential pocket, state $|X \rangle$, for $E_0=45$
MeV. We comment that, when including the inelastic channel $|2
\rangle$ but not $|Y \rangle$, the nucleus-nucleus potential
renormalization leads to the expected enhanced penetrability from
the inelastic channel coupling, compared to the purely elastic ($|1
\rangle$ plus $|X \rangle$) calculation. The decoherent dynamics due
only to environment $|X \rangle$ gives the (full curve). By
comparison, the calculation that also includes the GDR doorway-state
decay to $|Y\rangle$ leads to a suppression (dotted curve and arrow)
of the population of state $|X \rangle$. Additional irreversible
processes other than excitation of the GDR are also likely to
contribute to the deep-inelastic yield, such as complicated
multi-nucleon transfers \cite{Rehm}. To simulate these very simply,
the assumed state $|1 \rangle$ to $|2 \rangle$ coupling strength was 
increased. Figure\ \ref{Figure4} shows the dependence of the
calculated tunneling suppression on the assumed $\beta_1$ strength
for $E_0=45$, where we note
that larger $\beta_1$ result in both an increase in the strength and
the range of the coupling formfactor to the inelastic state $|2
\rangle$.

%\subsubsection{<A subsubsection>}
%%%%%%%%%%%%%%%%%%%%%%%%%%%%%%%%%%%%%%%%%%%%
%% Sample figure:
%%
%% The option [height=...] scales the picture to the given height,
%% without it it would be printed at its nominal size
%%%%%%%%%%%%%%%%%%%%%%%%%%%%%%%%%%%%%%%%%%%%

%\begin{figure}
%  \includegraphics[height=.3\textheight]{golfer}
%  \caption{Picture to fixed height}
%\end{figure}

%%%%%%%%%%%%%%%%%%%%%%%%%%%%%%%%%%%%%%%%%%%%
%% SAMPLE TABLE
%%
%% Shows the use of \tablehead and \tablenote
%% macros
%%%%%%%%%%%%%%%%%%%%%%%%%%%%%%%%%%%%%%%%%%%%

%\begin{table}
%\begin{tabular}{lrrrr}
%\hline
%  & \tablehead{1}{r}{b}{Single\\outlet}
%  & \tablehead{1}{r}{b}{Small\tablenote{2-9 retail outlets}\\multiple}
%  & \tablehead{1}{r}{b}{Large\\multiple}
%  & \tablehead{1}{r}{b}{Total}   \\
%\hline
%1982 & 98 & 129 & 620    & 847\\
%1987 & 138 & 176 & 1000  & 1314\\
%1991 & 173 & 248 & 1230  & 1651\\
%1998\tablenote{predicted} & 200 & 300 & 1500  & 2000\\
%\hline
%\end{tabular}
%\caption{Average turnover per shop: by type
%  of retail organisation}
%\label{tab:a}
%\end{table}

\section{SUMMARY}

We have reported on a quantum dynamical approach based on 
time-propagation of a coupled-channels density matrix. 
Both deep-inelastic processes and quantum tunneling (fusion) can be treated 
within this fully quantal framework. It describes 
the transition from pure state (coherent) to mixed state
(decoherent and dissipative) dynamics during a nuclear collision. 
The development provides a significant step towards an improved theoretical 
understanding of low-energy collision dynamics, as the calculations exhibit 
both quantum decoherence and energy dissipation. 
Effects of decoherence and dissipation on collision dynamics can be 
manifested at distances outside the fusion barrier radius, resulting in 
suppression of the quantum tunneling probability. This may have major 
implications for understanding current problems in near-barrier reaction 
dynamics \cite{Newton,Angeli,Nanda2}, including the sub-barrier fusion hindrance 
phenomenon \cite{Jiang}. More complete 
calculations and detailed consideration of other processes,
such as multi-nucleon or cluster transfer reactions, are required to confront 
measurements.

%%%%%%%%%%%%%%%%%%%%%%%%%%%%%%%%%%%%%%%%%%%%%%%%
%% BACKMATTER
%%%%%%%%%%%%%%%%%%%%%%%%%%%%%%%%%%%%%%%%%%%%%%%%

\begin{theacknowledgments}
Support from an ARC Discovery grant and the UK Science and
Technology Facilities Council (STFC) Grant No. EP/D003628/1 is
acknowledged. 
\end{theacknowledgments}




%\bibliography{sample}

%  \typeout{******************************************}
%  \typeout{** Please run "bibtex \jobname" to optain}
%  \typeout{** the bibliography and then re-run LaTeX}
%  \typeout{** twice to fix the references!}
%  \typeout{******************************************}

\end{document}